\documentstyle[11pt,newpasp,twoside,epsf]{article}
\def\emphasize#1{{\sl#1\/}}

\def\edcomment#1{\iffalse\marginpar{\raggedright\sl#1\/}\else\relax\fi}
\reversemarginpar

\begin{document}
\title{High Resolution X-Ray Spectroscopic Constraints on Cooling Flow
Models of Clusters of Galaxies and Gaseous Haloes Around Elliptical
Galaxies}
\author{S. M. Kahn, J. R. Peterson, F. B. S. Paerels, and H. Xu}
\affil{Columbia Astrophysics Laboratory, Columbia University, 550 W. 120th Street,
New York, NY  10027, United States}
\author{J. S. Kaastra, C. Ferrigno, T. Tamura, and J. Bleeker}
\affil{SRON Utrecht, Sorbonnelaan 2, 3584 CA Utrecht, The Netherlands}
\author{J. G. Jernigan}
\affil{Space Sciences Laboratory, University of California, Berkeley,
CA  94720, United States}

\begin{abstract}
In many clusters of galaxies, the cooling time at the core of the
intracluster medium is much less than the age of the system, suggesting that
the the gas should continually lose energy by radiation.  Simple thermodynamic
arguments then require that the expected ``cooling flow" should exhibit a
specific spectroscopic signature, characterized by a differential emission
measure distribution that is inversely proportional to the cooling function of
the plasma.  That prediction can be quantitatively tested for the first time
by the Reflection Grating Spectrometer (RGS) experiment on XMM-Newton, which
provides high resolution X-ray spectra, even for moderately extended sources
like clusters.  We present RGS data on 14 separate cooling flow clusters,
sampling a very wide range in mass deposition rate.  Surprisingly, in all cases we find a systematic deficit of low temperature emission relative to the
predictions of the cooling flow models.  However, we do see evidence for cooling
flow gas at temperatures just below the cluster background temperature, $T_0$,
roughly down to $T_0/2$.  These results are difficult to reconcile with most
of the possible explanations for the cooling flow problem that have been
proposed to date.  We also present RGS data on the massive elliptical galaxy
NGC 4636.  In this case, we detect evidence for resonance emission line 
scattering of high oscillator strength Fe L-shell emission lines within
the gaseous halo of the galaxy.  The detection of that effect leads to
very tight constraints on physical conditions within the halo.  However, 
here again, the expected signature of a cooling flow is not
detected, perhaps suggesting some fundamental uncertainty in our understanding
of radiative cooling in low density cosmic plasmas.
\end{abstract}

\section{Introduction}

The study of cooling flows in clusters of galaxies and the gaseous haloes of
ellipticals has been a controversial area of research for nearly 30 years.
As first emphasized independently by Cowie and Binney (1977), Fabian and
Nulsen (1977), and Matthews and Bregman (1978), the cooling time for gas at
the cores of these systems is often significantly less than the Hubble time.
This implies that the gas should continually lose energy by radiation.  A
pressure driven cooling flow is thought to develop, causing the gas in the 
densest regions to slowly accrete onto the central dominant galaxies.  The
net result is a steady deposition of cold matter in the core of the cluster
or massive galaxy.

General support for this picture came from some of the first imaging and
spectroscopic observations obtained with the \emphasize{Einstein}
Observatory.  Images of many clusters indicate sharp peaks in the surface
brightness distribution, as predicted by cooling flow models.  Where crude
spectrophotometry has been available, one typically sees a marked softening
of the spectrum in the central-most regions, again consistent with 
expectations.  Spectroscopically, the presence of the cooling flow should
manifest itself as an excess of emission (over isothermal models) at
the lowest energies, or via the presence of soft X-ray emission lines
inconsistent with the background gas temperature.  The latter effect
was detected with the Focal Plane Crystal Spectrometer on \emphasize{Einstein},
for the gaseous halo around M 87 (Canizares et al., 1979, 1982).

Nevertheless, subsequent observations led to refinements of the simplest
cooling flow models, and also pointed to some inconsistencies.  For example,
for a homogeneous cooling flow, the mass deposition rate should be even
more centrally peaked than is inferred from the X-ray brightness profiles
(Johnstone et al. 1992).  Typically, a mass deposition rate roughly
proportional to the radius is derived.  This was potentially explained by
Nulsen (1986) as a consequence of thermal instability.  Dense blobs form
locally everywhere in the intracluster medium where the cooling time is less than
the age of the system.  The result is that the cooling gas resides in a
\emphasize{multiphase} medium, distributed over a rather large volume.
However, this picture was criticized by Balbus and Soker (1989) and
others, who argued that in the presence of the gravitational potential,
the growth rates of the linear instability are too weak to allow the cooling
blobs to form.  

A potentially more serious problem is that the end products of cooling
are generally not seen in longer wavelength bands at the expected levels.  
For example, only a very small fraction of the cooling gas can form stars.
H I absorption measurements have found no evidence for cold condensed clouds
(O'Dea et al. 1998), and while CO emission has been detected (Edge 2001),
the amount of inferred molecular gas is a factor 10 below what is predicted
by X-ray cooling estimates.

Further, the first moderate resolution spectra obtained with the 
\emphasize{Einstein} Solid State Spectrometer and with ASCA did not provide
good quantitative fits to the simplest cooling flow models.  Although
cool gas is clearly present in the spectrum, there is a dearth of
emission at the lowest energies.  This was interpreted as evidence for 
cold absorbing material intermixed with the X-ray emitting gas (White
et al. 1991).  However, if cold gas is indeed present in the intracluster medium,
it remains unclear why it does not form stars.

\section{The Importance of High Resolution Soft X-ray Spectroscopy}

\begin{figure}
\plotone{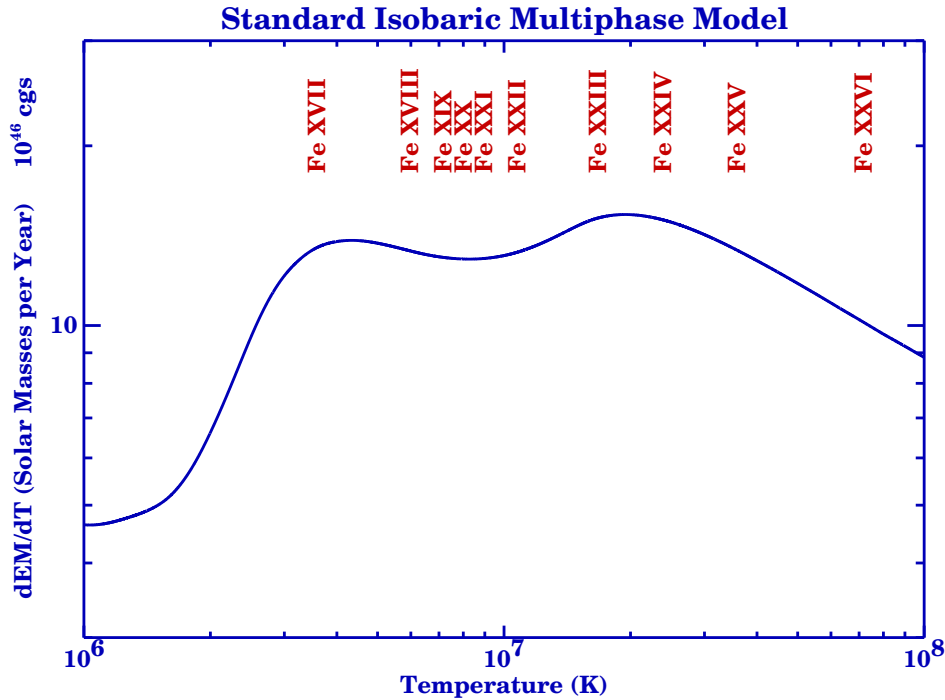}
\caption{The differential emission measure distribution of the isobaric multi-phase cooling flow model.  The peak of the emission of various Fe L ions are marked.  By resolving emission lines from these Fe L ions with the RGS, the model can be tested in considerable detail.}
\end{figure}

Independent of the details of the cooling flow process, the emitted
X-ray spectrum of the cooling gas can be robustly predicted using simple
thermodynamic arguments:  For an isobaric cooling flow, the differential
luminosity emitted in a small temperature range $dT$ is given by:

\begin{displaymath}
dL = \frac{5}{2} \frac{\dot{M}}{\mu m_p} kdT .
\end{displaymath}

$dL$ can also be expressed in terms of the cooling function of the 
plasma:

\begin{displaymath}
dL = \Lambda(T) dEM
\end{displaymath}

where $dEM = n^2 dV$ is the differential emission measure.  Thus:

\begin{equation}
\frac{dEM}{dT} = \frac{5}{2} \frac{\dot{M}}{\mu m_p} \frac{k}{\Lambda(T)} .
\end{equation}

The differential emission measure distribution predicted by this simple 
argument is displayed in Figure 1.  In calculating $\Lambda(T)$, we
have assumed one-third cosmic abundances, which is typical of intracluster
gas.  Marked in the figure are the respective temperatures where each
of the Fe L-shell ions becomes the dominant charge state.  As can be seen,
the Fe L complex very nicely samples the expected distribution.  Hence
by measuring the strengths of the various Fe L emission lines in a high resolution
soft X-ray spectrum, we can accurately constrain cooling flow models.

\section{Relevance of XMM-Newton}

The first experiment capable of unambiguously making such measurements is
the Reflection Grating Spectrometer (RGS) on the XMM-Newton Observatory.  
XMM-Newton incorporates three densely nested grazing incidence telescopes,
collectively providing unparalleled area for imaging and spectroscopy.  The
RGS consists of arrays of grazing incidence reflection gratings, mounted
immediately behind two of the three telescopes, which pick off roughly half
the light in the beams, and disperse it to sets of dedicated charge-coupled
device detectors (CCDs).  For each of the two arrays, the gratings are mounted 
on an inverted Rowland circle, which also includes the respective telescope
focus and the readout strip of CCDs.  The gratings are all identical, and they
are mounted at the same graze angle with respect to the incident ray
passing through each grating center.  This configuration produces nearly
stigmatic and aberration-free focussing at all wavelengths in the spectrum.
For a more complete description of this experiment, see den Herder et al.
(2002).  The other experiments on XMM-Newton are the European Photon
Imaging Camera (EPIC), arrays of CCDs mounted at the foci of the three
telescopes (see Turner et al. 2002 and Str\"{u}der et al. 2002), and the 
Optical Monitor (OM), an optical/ultraviolet telescope co-aligned with
with the X-ray instruments (Mason et al. 2002).

The RGS bandpass ($\lambda = 5$ to $38~{\rm \AA}$), was adjusted so as to optimally
sample the broad range of Fe L-shell transitions, as well as the 
K-shell features associated with C, N, O, Ne, Mg, and Si.  These are the 
most prominent emission lines expected for collisionally ionized gas,
emitting in the temperature regime $T = 10^6$ to $10^8 K$, characteristic
of the expected cooling flows in clusters and elliptical galaxies.  The effective
area over this band averages $\sim 120~{\rm cm}^2$, if we include both of the two
identical instruments.  For on-axis point sources, the spectral resolution is
$\Delta \lambda \approx 0.06$ \AA, varying only slowly as a function of wavelength.
Like the transmission grating experiments on the \emphasize{Chandra}
Observatory, the RGS is a slitless spectrometer, which means that the resolution
degrades for extended sources.  Nevertheless, since the dispersion of the
RGS gratings is very high, the degradation is fairly weak:

\begin{displaymath}
\Delta \lambda \approx 0.1 {\rm\AA} \times \Delta\theta
\end{displaymath}

\noindent
where $\Delta\theta$ is the source size in arc-minutes.  Thus, for only moderately
extended sources like clusters and ellipticals, we still obtain fairly
high resolution spectra.  (In contrast, the spectral resolution of the 
\emphasize{Chandra} experiments degrades by almost a factor of 100 for a
source extent of 1 arc-minute.)

\section{RGS Observations of Clusters and Groups}

\begin{figure}
\plotone{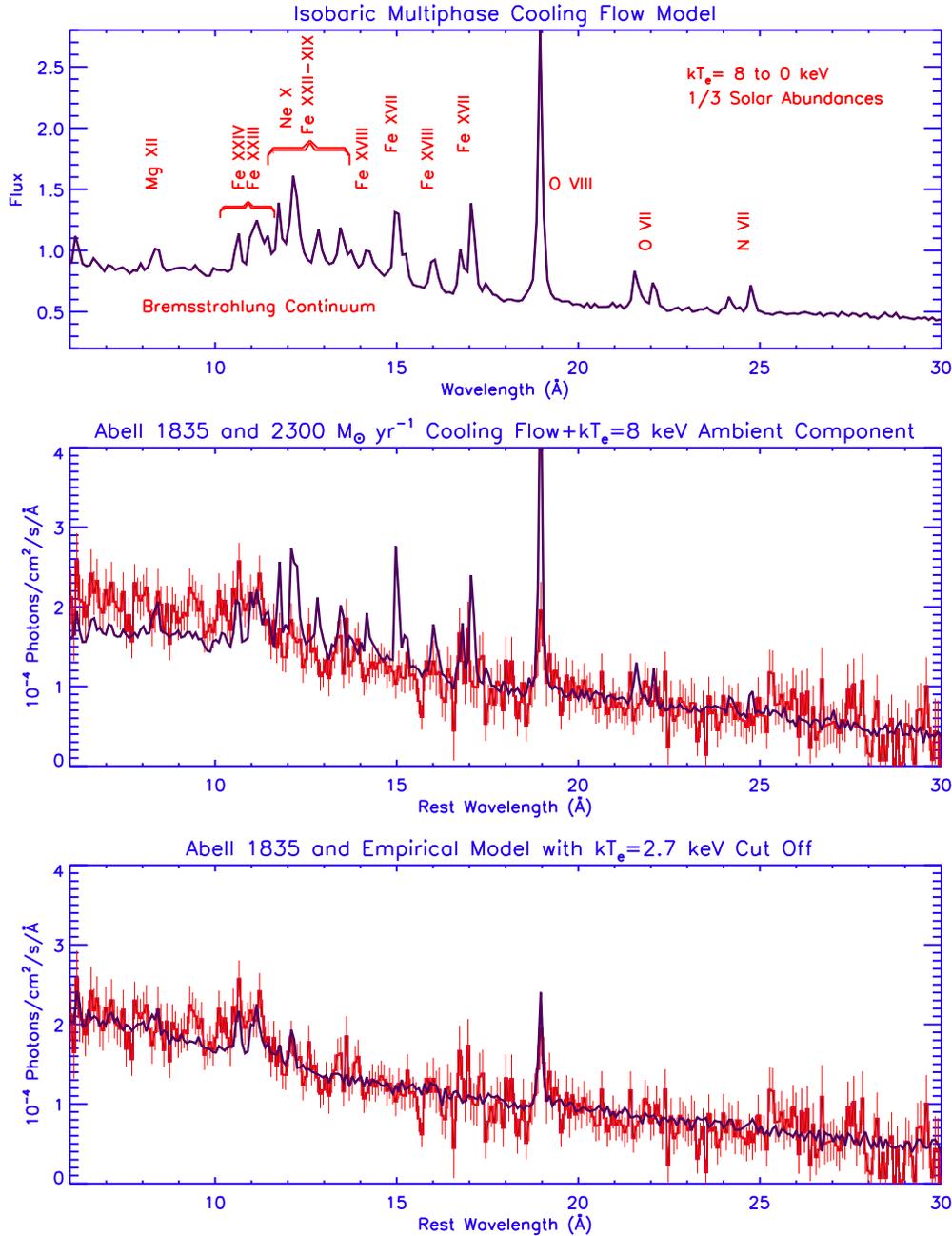}
\caption{The failure of the cooling-flow model in the massive cooling-flow, Abell 1835.  The top panel shows the spectral prediction for the cooling-flow model using the differential emission measure distribution in Figure 1 from the maximum temperature of 8 keV down to 0 keV.  The middle panel shows the comparison of the model (blue) with the data (red).  The model clearly overpredicts a number of Fe L emission lines.  The bottom panel shows the same model, but the emission below $\sim$ 3 keV is ignored.  Thus, there is evidence for cooling from 8 keV to 3 keV but no further.  See Peterson et al. (2001) for more information.}
\end{figure}

As a component of its guaranteed time observing program, the RGS instrument
consortium observed a sample of 14 compact clusters and groups of galaxies,
chosen to exploit the spectral sensitivity of the experiment (see Table 1).  For a complete discussion of the analysis of these data sets, see Peterson et al. (2002).
The sample  includes clusters and groups at a range of temperatures, from 1 to 10 keV.
All of these systems were expected to host cooling-flows, with inferred
mass deposition rates varying by three orders of magnitude ($\dot{M} \approx
1$ to $1000~M_{\odot}~{\rm yr}^{-1}$).  Surprisingly, it was apparent from even
the first of these observations, that our data were incompatible with the 
robust spectral predictions of the standard cooling flow model.  An example
is illustrated in Figure 2, which displays the derived spectra for the massive
cluster A 1835.  The intracluster medium in this system has a background 
temperature $kT_0 \approx 8$ keV, and an inferred mass deposition 
rate $\dot{M} \approx~2300 M_{\odot} {\rm yr}^{-1}$ (see Peterson et al. 2001 
for a more complete description of the analysis for this case.)  In the 
top panel of the figure, we show the simulated cooling flow spectrum that 
we expected for the parameters appropriate
to this source.  The middle panel shows the comparison with the data.  As can be
seen, the highest ionization lines expected (Fe XXIII, Fe XXIV) are indeed
detected in the data, however the model vastly overpredicts the intensity
of the lower ionization Fe L-shell lines, as well as the O VIII and O VII
K-shell transitions.  The discrepancy indicates a dearth of cooler gas.  In the
bottom panel, we show the comparison to the same cooling flow model, but now
with the predicted emission measure distribution truncated at $kT = 2.7$ keV,
with no contributions from gas at lower temperatures.  This somewhat arbitrary
modification of the model does appear to provide a fairly good description of
the data.

\begin{table}
\begin{center}
\begin{tabular}{lrr}
Cluster & Exposure (ks) & Redshift \\ \tableline
Abell 1835 & 36 & 0.2523 \\
Abell 665 & 20 & 0.1818 \\ 
Abell 1795 & 40 & 0.0622 \\
S\'{e}rsic 159-03 & 38 & 0.0580 \\
2A0335+096 & 26 & 0.0347 \\
Abell 4059 & 54 & 0.0460 \\
Abell 496 & 29 & 0.0328 \\
MKW 3s & 39 & 0.0442 \\
Abell 2052 & 33 & 0.0353 \\
Abell 262 & 36 & 0.0163 \\
Abell 1837 & 50 & 0.0372 \\
M87 & 42 & 0.0043 \\
NGC 533 & 48 & 0.0185 \\ \tableline
\end{tabular}
\caption{Clusters used in the RGS cooling-flow analysis.  See Peterson et al. (2002) for more details.}
\end{center}
\end{table}

Note that the data do \emphasize{not} indicate the complete absence of a cooling flow.
In particular, the detected Fe XXIII and XXIV lines are indicative of 
temperatures less than 4 keV, and cannot arise in the 8 keV gas that characterizes
most of the intracluster medium.  Interestingly, the intensities of these lines are
roughly at the right level for the mass deposition rate inferred from the
X-ray surface brightness.  It is as if the cooling does occur as predicted,
but the emergent luminosity somehow disappears from the soft X-ray band at temperatures
below $\sim$ half the background temperature of the cluster.

This same pattern is also found in all of the other clusters in our sample.
A subset of the measured spectra are shown in Figure 3.  Note that in all cases,
Fe XXIII and Fe XXIV lines are detected prominently, while emission lines
of lower ionization species (e.g. Fe XVII) are strongly suppressed.  While
O VIII Lyman-$\alpha$ ($\lambda = 18.97 \AA$) is always detected, its measured
intensity is also always far weaker than predicted by the cooling flow models for reasonable values of the oxygen abundance (O/Fe $>$ 0.2).

\section{Quantitative Spectral Fits}

\begin{figure}
\plotone{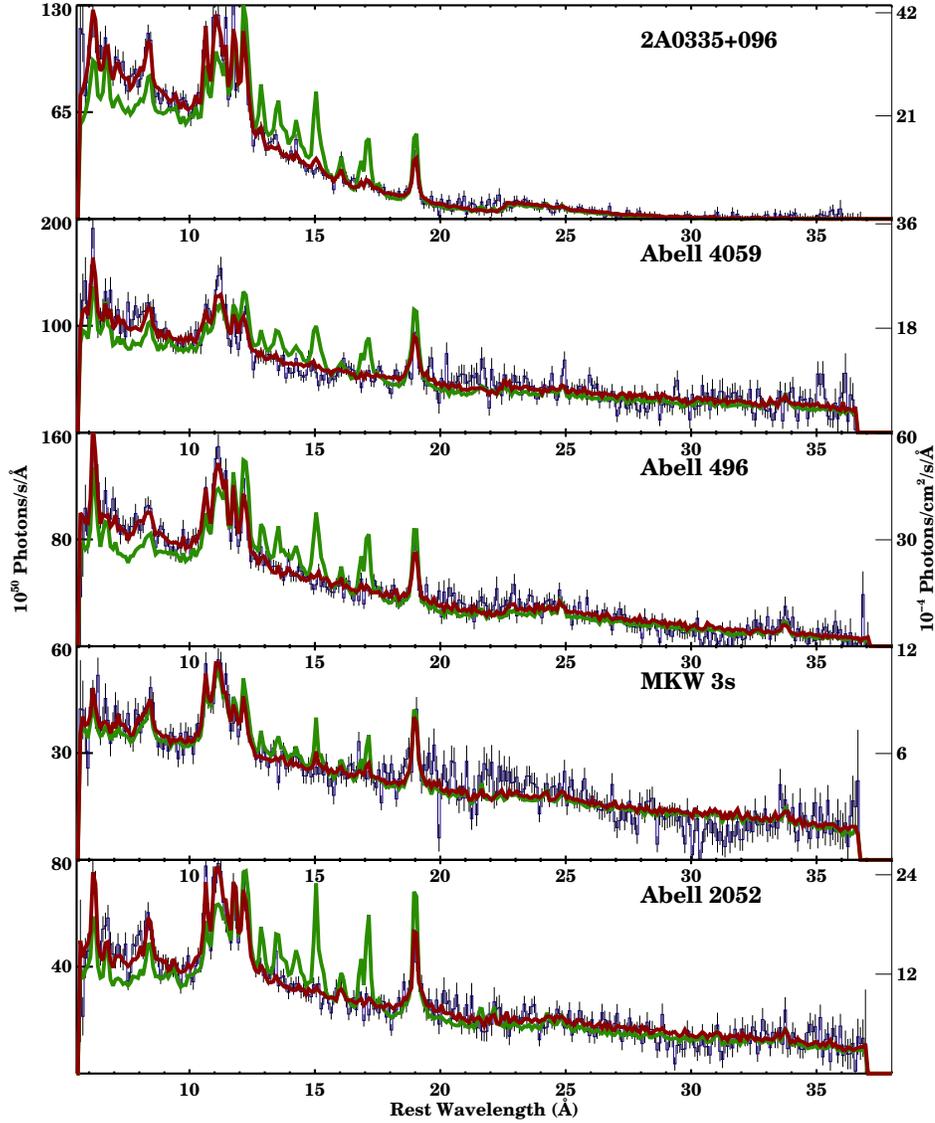}
\caption{The failure of the cooling-flow model in a set of clusters.  The blue points are the data, the green line is the cooling-flow model, and the red line is an empirical model where we have allowed the luminosity in various temperature bins to change.  All clusters show the largest deviation with the cooling-flow model for Fe XVII (15 and 17 \AA\/) and Fe XVIII (14.2 and 16 \AA\/).  For further details, see Peterson et al. (2002).}
\end{figure}

To derive more quantitative constraints on the discrepancies with the
cooling flow predictions, we have performed model fits to the data.  Since
clusters are extended sources, standard model fitting techniques cannot be easily
employed with RGS data.  Instead, we use a novel Monte Carlo technique developed 
expressly for this application (Peterson, Jernigan, and Kahn 2002).  The Monte Carlo 
correctly accounts for the off-axis behavior of the instrument response, and the effects
of arbitrary selection cuts and transformations of the data.

We generate photons within the cluster assuming a given spatial and spectral
model.  These are propagated through the instrument, yielding a distribution
in the dispersion coordinate, the cross-dispersion coordinate, and the CCD
pulse height.  After applying the same cuts as are applied to the real data,
the extracted spectrum from the Monte Carlo is compared to the data via
a $\chi^2$-statistic.  An iterative technique is used to optimize the fit
through the adjustment of the astrophysical parameters that specify the
spatial/spectral model.

\begin{figure}
\plottwo{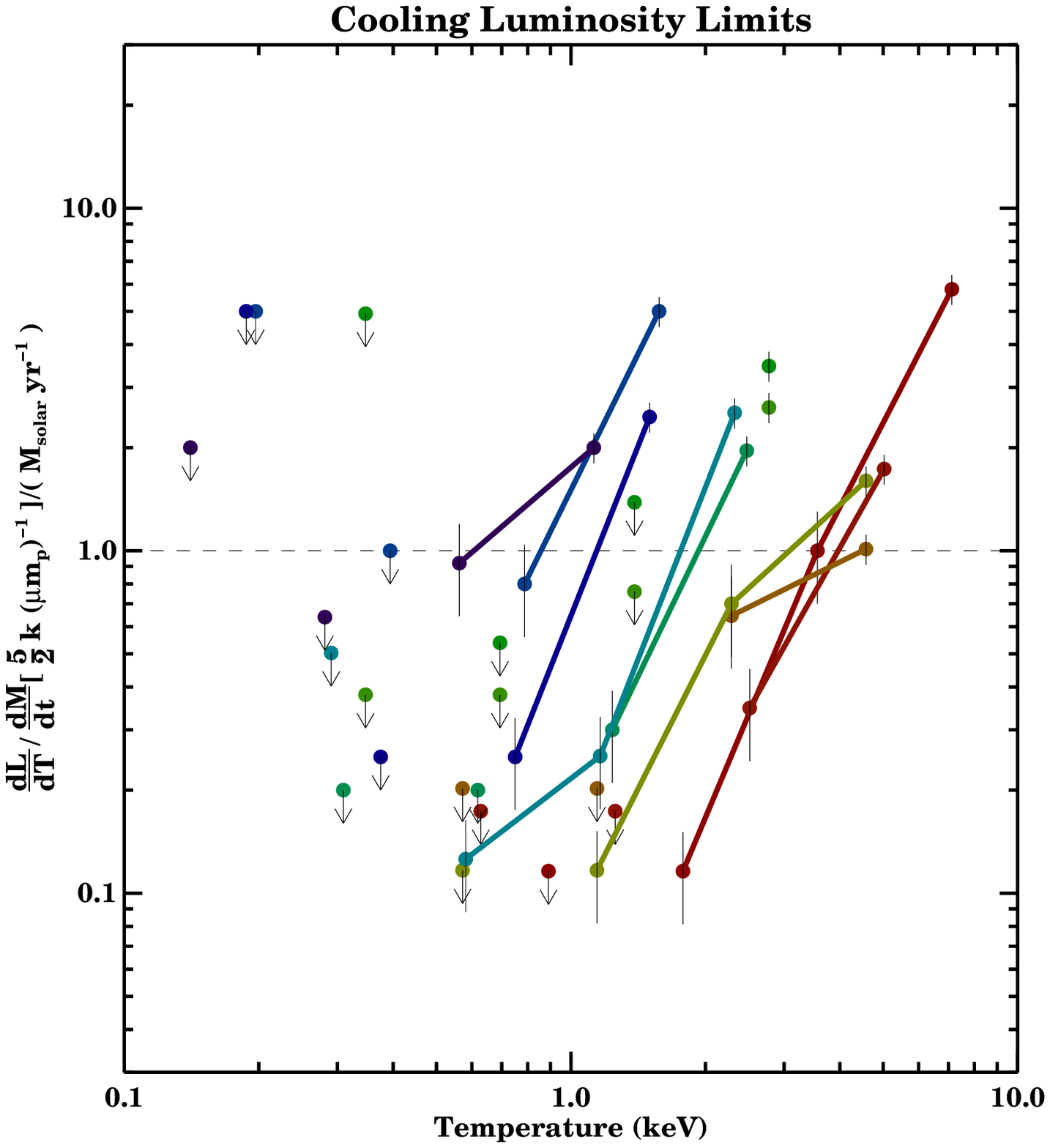}{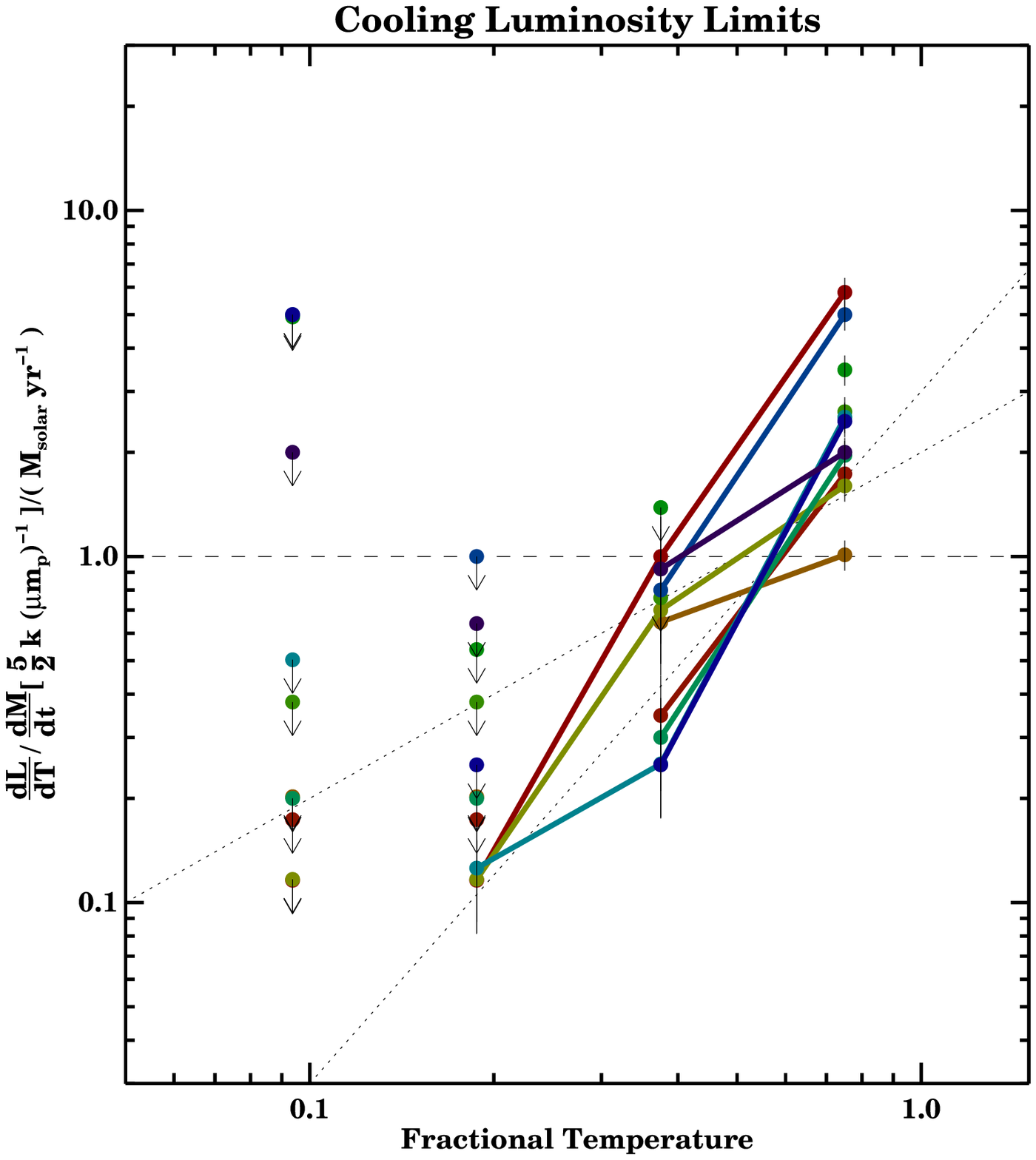}
\caption{Differential luminosity distribution of a sample of 14 clusters.  In the left panel the differential luminosity is plotted against the absolute temperature, and in the right panel the differential luminosity is plotted against the fraction of the background temperature.   The various clusters are plotted as different colors, and the points which are not upper limits are connected.  Both plots are normalized to the prediction from the isobaric cooling-flow model, which would predict all points to track the line $y$=1.  A number of detections and upper limits are well below that line, however, indicating the failure of the model at the lowest temperatures.  The right plot reveals a more systematic trend in the failure of the model.  See Peterson et al. (2002) for more details.}
\end{figure}

We use a simple $\beta$-model to characterize the distribution.  The parameters
are derived from the simultaneous EPIC images of the source.  For the spectrum,
we use the MEKAL (Mewe, Kaastra, and Liedahl 1995) model to generate the line
and continuum emissivities.  The shape of the spectrum is then determined by
the differential emission measure distribution.

Outside the ``break radius" determined from the EPIC image, we assume that
the gas is isothermal, and fit for the temperature and emission measure.  Inside,
we add in a cooling flow differential emission measure distribution, but we
allow the mass deposition rate, $\dot{M}$, to be normalized separately in 
different temperature bins:  ($\frac{1}{2}T_0$ to $T_0$), ($\frac{1}{4}T_0$
to $\frac{1}{2}T_0$), ($\frac{1}{8}T_0$ to $\frac{1}{4}T_0$), etc., where
$T_0$ is the background temperature.  This choice of binning is arbitrary,
but it works fairly well because the fractional ionization curves are roughly
equally spaced in the logarithm of the temperature.  The elemental abundances
are also left as free parameters, but are assumed to be spatially uniform.
The latter assumption does not have a big effect on the fits.

In general, the fits to the data with this model are quite good.  This can be
seen from Figure 3, where we show a representative sample of the data on individual
clusters with the best fit models overlayed.  The models appear to correctly
account for the relative strengths of the Fe L lines, O VIII, Ne X, and Mg
XII.  

In Figure 4, we plot the ratio of the derived differential luminosity for each
of the temperature bins to the differential luminosity predicted by the 
simple cooling flow model.  In Figure 4a, these points are plotted as a function
of absolute temperature, and in Figure 4b, as a function of fractional temperature
with respect to the background temperature for each individual cluster.  Different
clusters are represented by different color points, and the relative points
for a few particular clusters are connected by lines for clarity.  At the lowest
temperatures, we typically derive upper limits, as indicated, but in all cases,
we do derive clear detections of emission in at least two distinct temperature
bins.  

The horizontal dashed lines in these figures represent what would be expected
if the cooling flow model were correct.  While it is possible for the highest
temperature points to lie above the line (due to contributions from the background
temperature gas), one would expect the detections and/or upper limits to be
consistent with this line for all of the lower temperature points.  As can be
seen, in essentially all cases, at least some of the lower temperature points lie
well below the line, indicating a dearth of cooler gas.  When plotted as a
function of absolute temperature (Figure 4a), there is no other obvious pattern to the
data.  However, when plotted as a function of fractional temperature (Figure 4b),
it is clear that all of the clusters behave in roughly the same way:  there is
a clear ``drop-out'' of emission at temperatures below $\frac{1}{2}T_0$, where
$T_0$ is the background temperature.  

In fact, the various clusters appear
to be consistent with a differential emission measure distribution given by:

\begin{equation}
\frac{dEM}{dT} = \frac{5}{2} \frac{\dot{M}}{\mu m_p} \frac{k}{\Lambda(T)} \left( \alpha+1 \right)
\left( \frac{T}{T_0} \right)^{\alpha}
\end{equation}

with $\alpha \approx$ 1 to 2, as opposed to $\alpha = 0$, as required by
Equation (1).  This empirical relationship is indicated by the dashed sloping
lines in Figure 4b.  Thus, in addition to the model being inconsistent with the data at the lowest temperatures, the entire \emphasize{shape} of the differential emission measure distribution is inconsistent with the standard cooling-flow model.

\section{Possible Explanations for the Deficit of Soft Emission}

Since the initial reports of some of the results we have presented here
(Peterson et al. 2001, Kaastra et al. 2001, Tamura et al. 2001), a number of explanations
have been proposed to account for the observed deficit of soft X-ray emission.
Most of these work \emphasize{energetically}, i.e. they incorporate additional
heating or cooling mechanisms that can plausibly supply or remove energy at a
rate comparable to the missing soft X-ray luminosity.  However, it still unclear whether
any can successfully account for the systematic aspects of the observed
discrepancies, e.g. the fact that the drop-out of emission always seems to
occur at $\frac{1}{2}T_0$ over a wide range in $T_0$.  Below, we briefly discuss
a few of the ideas which have been suggested:

\emphasize{Heating}.  Heating by particle outflows associated with embedded active
galactic nuclei (Rosner \& Tucker 1989, Tabor \& Binney 1993, David et al. 2001), or
by subcluster mergers (Markevitch 2001), can supply sufficient energy deposition
to counteract the expected cooling.  In the context of the multiphase model, the
energy input must be distributed evenly everywhere over the thermally unstable region.
Otherwise local ``pockets" of gas may still cool and condense.  In addition, to
suppress cooling in a temperature regime that is a fixed fraction of the background
temperature, requires that the heating process be self-regulating, i.e. the heating
rate has to nearly balance the cooling rate for every cluster, over
a range of three orders of magnitude in cooling luminosity.  This seems rather
contrived, and, at the very least, such models require a fair degree of fine tuning 
to be considered viable.

\emphasize{Non-standard cooling}.  A more efficient means of cooling than radiation
in the soft X-ray regime could reduce the expected line emission.  Simply varying elemental
abundances does not help much, since iron dominates both the cooling and the observed
spectrum.  Mixing the hot gas with cooler embedded gas to reradiate the energy at
longer wavelengths (Begelman \& Fabian 1990, Fabian et al. 2001) is a possibility,
but it is by no means clear why this would always kick in at a fixed fraction of
the background temperature.  In addition, copious soft X-radiation can be produced
by charge transfer at hot/cold interfaces, so it might not be that easy to truly suppress
the soft X-ray emission line flux in such a scenario.

\emphasize{Electron conduction}.  Heat conduction
from hotter to colder phases could suppress further cooling as the dense condensations
first begin to form (Tucker \& Rosner 1983, Stewart et al. 1984, Bertschinger \& Meiksin
1986).  The presence of tangled magnetic fields in the plasma can strongly suppress
conduction down to levels where it becomes unimportant energetically (Chandran \& Cowley
1998), but this issue continues to be debated theoretically (Narayan \& Medvedev 2001).
Observationally, electron conduction has been observed to be suppressed by factors
$\sim$ 100 or more at cold fronts in clusters (Markevitch 2000), so the viability of
this explanation also remains suspect.

\clearpage
\begin{figure}
\plotfiddle{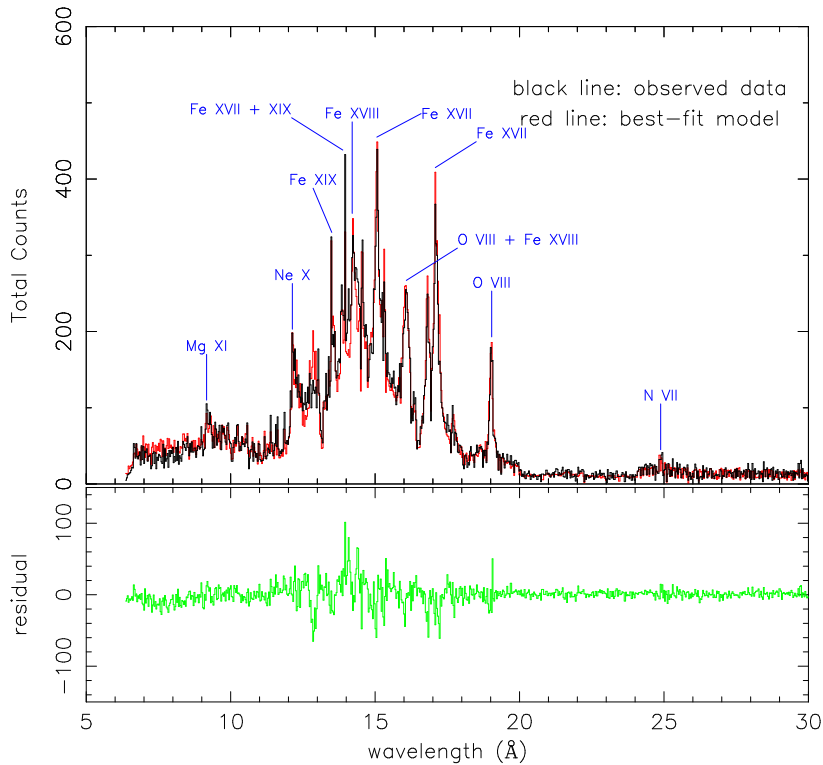}{4in}{-90}{60}{60}{-250}{320}
\caption{X-ray spectrum of the elliptical galaxy, NGC 4636.  The spectrum is dominated by emission lines from Fe XVII, Fe XVIII, and O VIII.  The red line is a spectral/spatial model using a Monte Carlo method to simulate the effects of resonant scattering.  See Xu et al. (2002) for further details.}
\end{figure}

\section{RGS Spectroscopy of the Elliptical Galaxy NGC 4636}

NGC 4636 is an E0-1 giant elliptical located in the skirt region of the Virgo cluster.
The gaseous halo associated with this galaxy has a total X-ray luminosity $L_X \approx
5.7 \times 10^{41} {\rm ergs~s}^{-1}$, making it one of the three brightest ellipticals in
the sky.  The X-ray temperature inferred from ASCA measurements is $kT \approx 0.76$ keV,
with inferred elemental abundances $\sim$ one third solar (Matsumoto et al. 1997).  At
this temperature, and the implied electron density, the gas at the core of the halo
should cool on timescales less than $10^8$ years, leading to a cooling flow with
a mass deposition rate $\dot{M} \sim 2~M_{\odot} {\rm yr}^{-1}$.

The RGS spectrum of NGC 4636 is shown in Figure 5.  As indicated, it is dominated
by very strong K-shell line emission from N, O, Ne, and Mg, as well a rich Fe L-shell
spectrum.  In contrast to the cluster spectra, here we see very strong emission
from the lower iron charge states, Fe XVII - XIX.  This is a consequence of the
much lower background temperature.  The red line in the figure is our best fit
thermal model for this spectrum, which involves a mean temperature $\sim 0.6$ keV
and elemental abundances ranging from 0.5 to 1 in solar units.  For further
details on the spectral fit, see Xu et al. 2002.

For this case, however, we can derive additional tight constraints on
physical conditions in the halo core by examining the spatial profiles of the data
in the cross-dispersion direction.  In particular, the RGS provides a nearly
stigmatic image of the source in the cross direction.  The emission line images
are observed to be extended, and the particular profile that we observe for each
line provides a clean indication of the distribution of the emission in that
line within the halo.

This is of interest for NGC 4636, because, for the inferred halo parameters, the
medium should be partially optically thick to resonance scattering for the high
oscillator strength Fe L transitions.  We observe this effect directly in the data.
In particular, the ratio of the Fe XVII $3s-2p$ lines at~17.1~\AA~to the Fe XVII
$3d-2p$ lines at 15.0~\AA\/ is found to be centrally peaked.  Both of these
sets of lines have similar emissivity dependences on temperature, so this cannot
be produced by a temperature gradient.  It is due to the fact that the high
oscillator strength 15.0~\AA\/ line scatters several times before exiting the halo, thereby broadening its spatial
distribution.

In fitting the data, we construct a Monte Carlo spatial/spectral model that is
self-consistently constrained to fit both the spectrum and the cross-dispersion 
emission line profiles.  The model is highly over-constrained.  For example, the
same density required to account for the line intensities, also yields the correct
Fe L line optical depths to reproduce the observed spatial profiles.  This can
only work at one value of the distance, and the distance we infer is consistent with
the known distance to the galaxy.

So for this particular case, there is essentially no ambiguity about the physical
conditions in the emitting plasma.  Nevertheless, the observed spectrum is still
dramatically inconsistent with the cooling flow predictions.  Specifically, at
the implied mass deposition rate, the Fe XVII line intensities should be a
factor 10 brighter than observed!  Similarly, the O VII line intensities are vastly
over-predicted.

Evidently, the multiphase cooling flow model also fails on the scale of an isolated
elliptical galaxy, and one without a currently active nucleus.  In our view, this single
observation provides one of the most stringent challenges to our understanding of
cooling in low density cosmic plasmas.

\section{Summary and Conclusions}

The first high spectral resolution X-ray observations of cooling flow clusters
continue to challenge conventional models for these systems.  Contrary to earlier
beliefs, the deficit of soft emission is not due to excess embedded absorbing
matter in the intracluster gas, but to a true lack of emission lines from the 
lowest temperature charge states.  We find this problem in all the systems we
have studied, from the most massive clusters to the gaseous haloes of individual
elliptical galaxies.

Empirically, our data appear consistent with the expression given by Equation (2),
where the emergent differential luminosity is proportional to the temperature
raised to a power $\alpha \sim 1$ to $2$, rather than being independent of 
temperature, as expected from simple thermodynamic arguments.   The solution to this 
puzzle may represent some fundamental new physics in the cooling of low density,
hot gas.  If so, it may have profound implications for models of galaxy and
structure formation in the early universe.

\acknowledgements
Analysis of XMM-Newton observations at Columbia University is supported by a
grant from NASA.  SRON Utrecht is supported financially by NWO, the Netherland
Organization for Scientific Research.

\end{document}